\documentclass{article}

\usepackage{enumitem}
\usepackage[table]{xcolor}
\usepackage{colortbl}

\PassOptionsToPackage{numbers, sort&compress}{natbib}
\usepackage{multirow}

\usepackage[preprint]{neurips_2025}

\usepackage[utf8]{inputenc} 
\usepackage[T1]{fontenc}    
\usepackage{hyperref}       
\usepackage{url}            
\usepackage{booktabs}       
\usepackage{amsfonts}       
\usepackage{nicefrac}       
\usepackage{microtype}      
\usepackage{xcolor}         
\usepackage[version=4]{mhchem}

\usepackage{amsmath}
\usepackage{amssymb}
\usepackage{mathtools}
\usepackage{amsthm}
\usepackage{xcolor}
\usepackage[most]{tcolorbox}
\usepackage[ruled,linesnumbered,vlined]{algorithm2e}
\usepackage{float}
\usepackage{wrapfig}

\usepackage{bm}
\usepackage{enumitem}

\newcommand{\minimize}[1]{\underset{{#1}}{\text{minimize}}}

\usepackage{esvect}

\makeatletter

\usepackage{letltxmacro}
\LetLtxMacro\orgvdots\vdots
\LetLtxMacro\orgddots\ddots

\usepackage{cleveref}

\theoremstyle{plain}

\theoremstyle{definition}

\theoremstyle{remark}

\newcommand{\BiTe}{\ce{Bi2Te3}}

\title{Constrained Diffusion for Accelerated Structure Relaxation of Inorganic Solids with Point Defects}

\author{%
  Jingyi Cui\\
  University of Virginia\\
  \texttt{\href{mailto:cau8rc@virginia.edu}{cau8rc@virginia.edu}}
  \And
  Jacob K. Christopher\\
  University of Virginia\\
  \texttt{\href{mailto:csk4sr@virginia.edu}{csk4sr@virginia.edu}}
  \And
  Ankita Biswas\\
  University of Virginia\\
  \texttt{\href{mailto:ab8ky@virginia.edu}{ab8ky@virginia.edu}}
  \And
  Prasanna V. Balachandran\\
  University of Virginia\\
  \texttt{\href{mailto:pvb5e@virginia.edu}{pvb5e@virginia.edu}}
  \And
  Ferdinando Fioretto\\ 
  University of Virginia\\
  \texttt{\href{mailto:fioretto@virginia.edu}{fioretto@virginia.edu}}
}

\begin{document}

\maketitle

\begin{abstract}
Point defects affect material properties by altering electronic states and modifying local bonding environments. However, high-throughput first-principles simulations of point defects are costly due to large simulation cells and complex energy landscapes.
To this end, we propose a generative framework for simulating point defects, overcoming the limits of costly first-principles simulators.
By leveraging a primal-dual algorithm, we introduce a constraint-aware diffusion model which outperforms existing constrained diffusion approaches in this domain. Across six defect configuration settings for \ce{Bi2Te3}, the proposed approach provides state-of-the-art performance generating physically grounded structures.
\end{abstract}

\vspace{-8pt}
\section{Introduction}
\label{sec:intro}
\vspace{-8pt}

Point defects play an important role in determining the properties of crystalline materials \cite{li2010high, su2017multi, goyal2017computational, mosquera2023identifying}. Recent work on additive manufacturing (AM) of thermoelectric materials reveals a more complex relationship between processing conditions and point defects than in traditional methods \cite{bi2024additive, welch2021nano, oztan2022additive}. They are exploited to tailor the structural, electronic, and thermal properties of advanced materials for technological applications.

Layered chalcogenides, a promising material class, are highly susceptible to point defects due to their mixed bonding with weak interlayer van der Waals interactions and strong intralayer covalent bonds, which critically influence electronic and phononic transport properties \cite{cain2016emerging, romanenko2021review}. A well-known example is bismuth telluride (\BiTe), which has been widely studied for its high thermoelectric efficiency in the low-temperature regime \cite{pan2015electrical, shen2011texture, pan2016thermoelectric, tang2019bi}. The performance of \BiTe~is strongly governed by defect concentration and configuration, and preliminary AM work has shown that laser processing parameters can change thermoelectric properties to optimize efficiency and carrier transport. \cite{welch2021nano, oztan2022additive}.

Point defects are hard to probe directly and are typically identified indirectly via combined techniques. Traditionally, density functional theory (DFT) have been used to complement experimental methods for defect identification \cite{RevModPhys.86.253} which is computationally expensive, making high-throughput studies computationally prohibitive. This situation highlights the need for alternative approaches, such as surrogate models, to supplement the DFT calculations. In related domains, deep learning methods have successfully overcome these barriers \cite{chun2020deep}. Generative models enable digital twins for defect prediction, with diffusion models achieving state-of-the-art material generation and inverse design \cite{dong2024generative, park2024inverse, takahara2024generative}.
Yet, these models excel at generating realistic data, struggle with strict physical constraints. Without reliability, designs risk impracticality and may hinder transition to production.

For other scientific applications, these challenges have been addressed through physics-aware generative processes. Simple constraints can be injected into sampling \cite{christopher2024constrained, cheng2024gradient, utkarsh2025physics}, but complex ones  often require integrating costly simulators with high overhead \cite{yuan2023physdiff, zampini2025training}. In this case, runtime demands may limit utility, making rejection sampling more efficient. For physics-informed generation of materials like \BiTe, first-principle simulators \cite{giannozzi2009quantum} are challenging, as constraints lack closed forms and evaluations scale exponentially with system size.

\textbf{Contributions.}
To address these existing challenges, this paper makes the following contributions: 
\textbf{(1)} It extends previous constrained generation approaches to handle complex constraints modeled by tractable functions and neural surrogates, leveraging a primal-dual algorithm.
\textbf{(2)} It provides the first study which explores constrained diffusion to supplement the DFT-based point defect simulation of a thermoelectric material (\BiTe).
\textbf{(3)} Through rigorous evaluation across six defect configurations of interest, we provide state-of-the-art performance for generative modeling of physically realistic predictions of layered chalcogenides.

\begin{figure}
    \centering
    \includegraphics[width=0.8\linewidth]{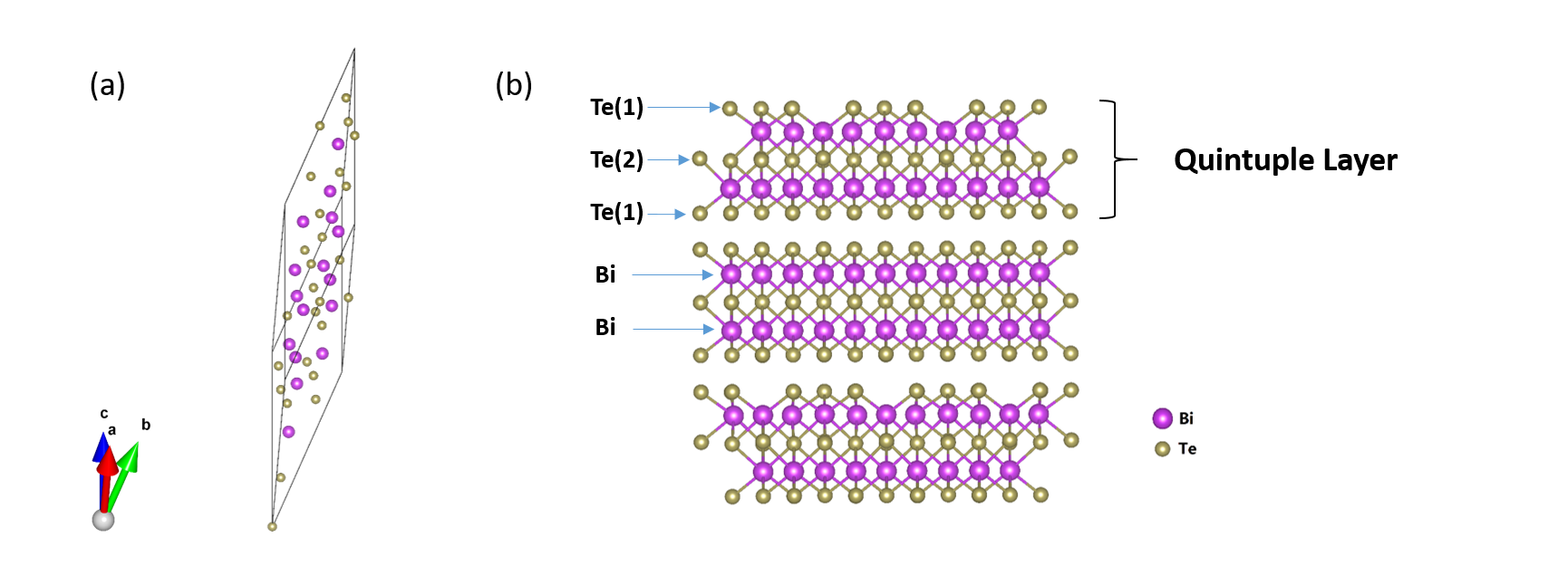}
    \caption{\textbf{(a)} A 2$\times$2$\times$2 supercell of \ce{Bi2Te3} for DFT simulations. \textbf{(b)} Another supercell representation, highlighting quintuple layer, structure Te(1)–Bi–Te(2)–Bi–Te(1) with vacancy defect.}
    \label{fig:bite_illustration}
    \vspace{-15pt}
\end{figure}

\vspace{-8pt}
\section{Preliminaries}
\label{sec:prelim}
\vspace{-8pt}

\textbf{Score-based Diffusion Models} \cite{song2019generative, song2020score},
and equivalently denoising diffusion probabilistic models \cite{ho2020denoising, park2024random}, model data distributions by introducing a forward noising process and training a neural network to approximate its reverse. In the \emph{forward process}, a clean sample $\bm{x}_0 \sim p_\text{data}(\bm{x}_0)$ is gradually perturbed through a sequence of intermediate distributions $\{p_t(\bm{x}_t)\}_{t=0}^T$. This is achieved using a Gaussian noise kernel with a predefined variance schedule $\bar{\alpha}_t$, which increases with $t$. As $t \rightarrow T$, the distribution of $\bm{x}_t$ converges to a standard Gaussian, such that $p_T(\bm{x}_T) \approx \mathcal{N}(0, I)$.
This process facilitates the training of a score network \(\bm{s}_\theta (\bm{x}_t, t)\), which learns the score function \(\bm{s}_\theta (\bm{x}_t, t) \approx \nabla_{\bm{x}_t} \log p_t (\bm{x}_t |\bm{x}_0)\) by error in the predicted score estimate:
\begin{align}
    \min_\theta \underset{t \sim [T, 1], \; \bm{x}_0 \sim p_\text{data}}{\mathbb{E}} \bigl[ \| \bm{s}_\theta(\bm{x}_t, t) - \nabla_{\bm{x}_t} \log p_t (\bm{x}_t |\bm{x}_0) \|_2^2\bigr]
\end{align}
The trained score network \(\bm{s}_\theta (\bm{x}_t, t)\) is then used in the \emph{reverse process} to iteratively reconstruct data samples from the training distribution \(p_\text{data}\). At each timestep $t$, the score function is applied to the reverse process, transitioning Gaussian noise \(\bm{x}_T \sim \mathcal{N}(0, I)\) to high fidelity samples.

\textbf{Projected Diffusion Models} \cite{christopher2024constrained}
build from Langevin Dynamics employed by score-based diffusion models, interpreting the sampling process as an optimization with respect to a series of intermediate probability density functions (Equation \eqref{eq:constrained-diffusion}).
In doing so, the formulation enabled a natural extension to constrained sampling problems by enforcing constraint adherence on the sample:
\begin{subequations}
\label{eq:constrained_optimization}
    \begin{align}
        \label{eq:constrained-diffusion}
        \minimize{\bm{x}_{T}, \ldots, \bm{x}_0} &\;
        \sum_{t = T, \ldots, 0}- \log p_t(\bm{x}_{t}|\bm{x}_0) \\
        \label{eq:constrained-diffusion-constr}
        \textrm{s.t.:}  &\quad \bm{x}_{T}, \ldots, \bm{x}_0 \in \mathbf{C}.
    \end{align}
\end{subequations}
Constraints can then be enforced by extending the reverse process with Projected Langevin Dynamics, such each update step is projected to the nearest feasible point by \(\mathcal{P}_\mathbf{C}(\bm{x}) = \underset{\bm{y}\in\mathbf{C}}{\arg\min} \|\bm{y} - \bm{x}\|_2^2\):
\begin{align}
    \label{eq:projection_update}
    \bm{x}_t^{(i+1)} = \mathcal{P}_\mathbf{C} \bigl(\bm{x}_t^{(i)} + \gamma_t \bm{s}_\theta (\bm{x}_t^{(i)}, t) + \sqrt{2 \gamma_t}\epsilon \bigr)
\end{align}
where \(\gamma_t\) is the step size decreasing with $t$, and \(\epsilon\) is Gaussian noise.

\vspace{-8pt}
\section{Constrained Diffusion for Point Defect Generation}
\label{sec:tdcd}
\vspace{-8pt}

Generating relaxed atomic structures with targeted defects in \ce{Bi2Te3} is a challenging problem, as the combinatorial diversity of defect placements creates a vast configuration space where DFT becomes computationally prohibitive.
To address this, we propose a \emph{generative diffusion framework} that samples \ce{Bi2Te3} structures under prescribed stoichiometries. 
Given target counts of Bi and Te atoms in a supercell, as shown in Figure \ref{fig:bite_illustration}, the model generates diverse arrangements that explore physically meaningful defects while retaining diffusion sampling's inherent randomness. 
However, diffusion alone does not ensure physical plausibility, motivating the adoption of three constraint types.

\begin{enumerate}[left=0pt,nosep,topsep=-2pt]
\item \emph{Geometric Constraints.} Ensure physically-meaningful interatomic distances and periodic boundary conditions. Physical validity requires a lower bound on interatomic distances to prevent atomic collapse. For consistency with DFT simulations, generated structures must also satisfy periodic boundary constraints to align with reference distributions. 
Thus,
\begin{equation}
\|r_i - r_j\| \;\ge\; d_{\min}, \;\; \forall i \neq j;
\qquad f_i \in [0,1]^3,
\end{equation}
where $d_{\min}$ is the minimum allowable interatomic distance between arbitrary atoms \(r_i\) and \(r_j\), 
and $f_i$ is the fractional coordinate constrained to $[0,1]^3$ to enforce periodic boundary conditions.

\item \emph{Distributional Constraints.} Align radial distributions (within 5\r{A}) with reference data; This reflects structural patterns, such as relative intensities of coordination shells. 
This yeilds,
\begin{equation}
D_{[0,R_c]}\!\left(p_{\text{gen}}, \, p_{\text{ref}}\right) \le \varepsilon,
\end{equation}
where $D_{[0,R_c]}$ measures discrepancy between distributions over distances in $[0,R_c]$, $p$ denotes the empirical distribution of pairwise distances in generated and reference structures. 

\item \emph{Force Minimization.} Stable structures are defined by low total forces, rendering force minimization critical for physical plausibility. As shown in Section \ref{sec:exp}, stochastic generations often yield high forces. We therefore introduce a \emph{soft} force-minimization constraint, adding a $\|\mathrm{force}(r)\|^2$ prediction term from a surrogate \cite{batatia2022mace} to the objective \eqref{eq:constrained-diffusion}, optimized at sampling.
\end{enumerate}

\textbf{Terminal Set Constraints.}
Projection-based constraint enforcement can be very effective in many applications. Unlike post-processing, these techniques yield higher-fidelity outputs to the learned distribution, as the score function refines samples displaced from high-density regions. Yet, projected diffusion models rely on a stringent assumption: \emph{Constraints can be meaningfully imposed on any $\bm{x}_t \sim p_t(\bm{x}_t)$ for an arbitrary $t$} \cite{christopher2024constrained}.

Often, it may be reasonable to assume that constraints can be modeled at intermediate states (e.g., path planning, where even noisy samples form valid trajectories) \cite{liang2025simultaneous}. However, this assumption does not hold in many scientific applications such as ours, where complex dynamics require surrogates that are unreliable on noisy samples.
This challenge, studied in training-free guidance \cite{ye2024tfg, yu2023freedom, cornwalltraining}, can lead to ``misaligned gradients,'' only matching the true value as \(t \rightarrow 0\), leading to compounding errors \cite{shen2024understanding, vaeth2024gradcheck}.
Alternatively, finetuning on noisy data is often ineffective: \textbf{(1)} Labels are not available for our constraints due to simulators requirements \cite{giannozzi2009quantum}, which are not only prohibitive due to runtime but also as the DFT process fails on increasingly noisy samples. \textbf{(2)} Even when labels are available, gradient inaccuracies remain a concern \cite{kawar2022enhancing}.

Hence, designing a constraint evaluation surrogate robust to high noise levels is prohibitive for complicated constraint sets. 
A more promising strategy is to relax the intermediate feasibility assumption.
To facilitate this, our framework treats the objective in Equation \eqref{eq:constrained_optimization} as a \emph{multi-stage optimization problem}.
While Equation \eqref{eq:constrained-diffusion-constr} enforces $\bm{x}_t \in \mathbf{C}$, from a multi-stage perspective only the terminal state $\bm{x}_0$ must be feasible.
Thus, intermediate constraints serve as a conservative proxy for final feasibility, while earlier states may remain unconstrained if the process yields \(\bm{x}_0 \in \mathbf{C}\).

Hence, we impose the constraints only on the final minimization of \(- \log p_\text{data}(\bm{x}_0)\).
This can then be interpreted as terminal set constraints drawn from Model Predictive Control theory \cite{rawlings2020model}, where feasibility is often only enforced on final decision variables. 
This removes dependence on intermediate feasibility assumptions as constraints are only imposed on clean samples within \(p_\text{data}\).

\textbf{Primal-Dual Projection Algorithm.}
As neural surrogates provide differentiable loss functions, it is most reasonable to solve the projection mapping using a gradient-based approach. To facilitate this, we adopt the Augmented Lagrangian representation of the constraint problem following \citeauthor{liang2025simultaneous}, incorporating constraints into the objective through Lagrange multipliers \cite{boyd2004convex}, and yielding:
\begin{align}
    \mathcal{L}(\bm{y},\bm{x}_0^{(i)}; \lambda, \mu) := \|\bm{y} - \bm{x}_0^{(i)}\|^2_2 + \sum_j^N \lambda_j c_j(\bm{y}) + \sum_k^N \frac{\mu_k}{2} c_k(\bm{y})^2
\end{align}
where \(\lambda = \{\lambda_1,\ldots,\lambda_N\} \) and \(\mu = \{\mu_1,\ldots,\mu_N\} \) are dual variables associated with $N$ constraint violation functions \(c = \{c_1, \ldots, c_N\}\) and \(\bm{y}\) is the solution optimized. Ensuring the strongest relaxation of this projection, the Lagrangian dual is used to optimize the dual variables \cite{fioretto2020lagrangian}:
\begin{align}
    \underset{\lambda, \;\mu}{\arg \max} \; \Bigl( \underset{\bm{y}}{\arg\min} \bigl( \mathcal{L}(\bm{y},\bm{x}_0^{(i)}; \lambda, \mu) \bigr) \Bigr)
\end{align}
This converges to satisfy the distance and distributional constraints while simultaneously minimizing the force and remaining close to the original input. Following Equation \eqref{eq:projection_update}, we begin applying this as our projection operator at $t = 0$. More details are provided in Appendix \ref{app:alm}.

\begin{table}[t]
\centering
\resizebox{0.6\columnwidth}{!}{
\begin{tabular}{llccc}
\toprule
Method & Config & RMSD $\downarrow$ & RDF $\downarrow$ & Force (eV/\r{A}) $\downarrow$  \\
\midrule
\multirow{2}{*}{Conditional DM} 
 & 16Bi + 21Te & $1.86 \pm 0.33$ & $65.85 \pm 19.75$ & $1.85 \times 10^{9}$ \\
 & 16Bi + 22Te & $1.93 \pm 0.36$ & $82.11 \pm 29.54$ & $3.49 \times 10^{9}$ \\
\midrule
\multirow{2}{*}{Projected DM} 
 & 16Bi + 21Te & $2.54 \pm 0.55$ & $1.11 \pm 0.77$ & $8.35 \times 10^{2}$ \\
 & 16Bi + 22Te & $1.41 \pm 0.83$ & $0.78 \pm 0.55$ & $7.20 \times 10^{1}$ \\
\midrule
\multirow{2}{*}{Post-proc DM} 
 & 16Bi + 21Te & $2.30 \pm 0.40$ & $27.25 \pm 28.30$ & $8.45 \times 10^{6}$ \\
 & 16Bi + 22Te & $1.14 \pm 0.65$ & $21.82 \pm 28.09$ & $9.29 \times 10^{6}$ \\
\midrule
\rowcolor{gray!12}
\textbf{Ours}
 & 16Bi + 21Te & $\mathbf{0.90} \pm 0.38$ & $\mathbf{0.30} \pm 0.11$ & $\mathbf{7.88 \times 10^{-2}}$ \\
\rowcolor{gray!12}
 & 16Bi + 22Te & $\mathbf{1.02} \pm 0.70$ & $\mathbf{0.35} \pm 0.18$ & $\mathbf{1.44 \times 10^{-1}}$ \\
\bottomrule
\end{tabular}
}
\vspace{5pt}
\caption{Empirical comparison across stoichiometric configurations. Extended table in Appendix \ref{app:setup}.}
\label{tab:methods}
\vspace{-20pt}
\end{table}


\vspace{-8pt}
\section{Experiments}
\label{sec:exp}
\vspace{-8pt}
\textbf{Baselines.}
We compare our approach against three representative baselines: \textbf{(1)} conditional diffusion models \cite{ho2022classifier}, where constraints are incorporated during training and sampling to bias the denoising process toward feasible regions. 
\textbf{(2)} post-processing optimization \cite{giannone2023aligning}, where the final output of an unconstrained diffusion model is projected onto the constraint set in a single step. 
\textbf{(3)} projected diffusion models \cite{christopher2024constrained}, which enforce constraints at every denoising step. This represents the current state-of-the-art but relies on meaningful constraint evaluation under noisy configurations.

\textbf{Evaluation.}
We target four categories of defects that frequently occur in \ce{Bi2Te3} structure systems: Te vacancies, Bi vacancies, Bi on Te anti-site, and Te on Bi anti-site. 
Additional details on the problem setup and results are provided in Appendix \ref{app:setup}.
Our evaluation consists of the following metrics:
\begin{itemize}[leftmargin=10pt,labelindent=0pt,labelsep=0.2em,itemsep=0pt,topsep=0pt]
\item \textbf{RMSD (Root Mean Square Deviation)}:
Computes positional deviation between generations and closest reference structures by matching atoms via a Hungarian matching algorithm.
\item \textbf{RDF (Radial Distribution Function)}: Compares radial distributions for all generations within 5\r{A} to assess global geometric consistency. This indicates the accuracy of positions and heights of nearest neighbor shell peaks under a common defect. 

\item \textbf{Force (eV/\r{A})}: Reports the total force as computed by a neural surrogate \cite{batatia2022mace}. Lower force indicates generations closer to stable states and more suitable for downstream simulations.

\end{itemize}

\textbf{Interpretation.}
As shown in Table~\ref{tab:methods}, our method achieves substantially lower deviation from reference structures and reduced total forces.
Existing approaches observe key limitations: \textbf{(1)} conditional diffusion preserves structural characteristics but allows overlapping atoms, undermining physical realism and leading to extremely high total force; \textbf{(2)} post-processing, while improving over the conditional model, struggles to impose the highly nonconvex constraints in a single step, as the final output often deviates from the constraint set significantly; \textbf{(3)} projected diffusion provides the strongest baseline, but misaligned gradients at higher noise results in higher RDF deviation.
In contrast, our method preserves diffusion dynamics while ensuring feasibility. By applying constraints on the final Langevin optimization, gradient inaccuracies are avoided (unlike projected diffusion models), while providing a sufficient number of constrained diffusion steps to ensure feasibility (unlike post-processing schemes). We achieve the lowest RMSD in nearly all settings, the highest RDF similarity (by more than 2x), and lower total forces by several orders of magnitude.

\vspace{-8pt}
\section{Conclusion}
\label{sec:conclusion}
\vspace{-8pt}

Motivated by the real-world importance of thermoelectric defect structure modeling, this paper addresses existing barriers by integrating constraints into the sampling process through tractable functions and neural surrogate models. Building upon a primal–dual projection algorithm, this work provides effectively enforces feasibility while addressing challenges with misaligned gradients. Across six representative defect configurations in \ce{Bi2Te3}, our approach provides state-of-the-art performance in producing physically realistic structures, highlighting the significance of our framework for complex material systems.

\vspace{-8pt}
\section*{Acknowledgements}
\vspace{-4pt}

This research is partially supported by NSF awards 2533631, 2401285 and by DARPA under Contract No. \#HR0011252E005. 
The authors acknowledge the Research Computing at the University of Virginia. 
Any opinions, findings, conclusions, or recommendations expressed in this material are those of the authors only.

\vspace{-4pt}

\bibliography{bib}

\begin{thebibliography}{45}
\providecommand{\natexlab}[1]{#1}
\providecommand{\url}[1]{\texttt{#1}}
\expandafter\ifx\csname urlstyle\endcsname\relax
  \providecommand{\doi}[1]{doi: #1}\else
  \providecommand{\doi}{doi: \begingroup \urlstyle{rm}\Url}\fi

\bibitem[Li et~al.(2010)Li, Liu, Zhao, and Zhou]{li2010high}
Jing-Feng Li, Wei-Shu Liu, Li-Dong Zhao, and Min Zhou.
\newblock High-performance nanostructured thermoelectric materials.
\newblock \emph{NPG Asia Materials}, 2\penalty0 (4):\penalty0 152--158, 2010.

\bibitem[Su et~al.(2017)Su, Wei, Li, Liu, Yan, Li, Su, Xie, Zhao, Zhai,
  et~al.]{su2017multi}
Xianli Su, Ping Wei, Han Li, Wei Liu, Yonggao Yan, Peng Li, Chuqi Su, Changjun
  Xie, Wenyu Zhao, Pengcheng Zhai, et~al.
\newblock Multi-scale microstructural thermoelectric materials: transport
  behavior, non-equilibrium preparation, and applications.
\newblock \emph{Advanced Materials}, 29\penalty0 (20):\penalty0 1602013, 2017.

\bibitem[Goyal et~al.(2017)Goyal, Gorai, Peng, Lany, and
  Stevanovi{\'c}]{goyal2017computational}
Anuj Goyal, Prashun Gorai, Haowei Peng, Stephan Lany, and Vladan
  Stevanovi{\'c}.
\newblock A computational framework for automation of point defect
  calculations.
\newblock \emph{Computational Materials Science}, 130:\penalty0 1--9, 2017.

\bibitem[Mosquera-Lois et~al.(2023)Mosquera-Lois, Kavanagh, Walsh, and
  Scanlon]{mosquera2023identifying}
Irea Mosquera-Lois, Se{\'a}n~R Kavanagh, Aron Walsh, and David~O Scanlon.
\newblock Identifying the ground state structures of point defects in solids.
\newblock \emph{npj Computational Materials}, 9\penalty0 (1):\penalty0 25,
  2023.

\bibitem[Bi et~al.(2024)Bi, Liu, Li, Li, Yang, Starostenkov, and
  Dong]{bi2024additive}
Jiang Bi, Zeqi Liu, Bo~Li, Shide Li, Zhuoyun Yang, Mikhail~Dmitrievich
  Starostenkov, and Guojiang Dong.
\newblock Additive manufacturing of thermoelectric materials: materials,
  synthesis and manufacturing: a review.
\newblock \emph{Journal of Materials Science}, 59\penalty0 (2):\penalty0
  359--381, 2024.

\bibitem[Welch et~al.(2021)Welch, Hobbis, Birnbaum, Nolas, and
  LeBlanc]{welch2021nano}
Ryan Welch, Dean Hobbis, Andrew~J Birnbaum, George Nolas, and Saniya LeBlanc.
\newblock Nano-and micro-structures formed during laser processing of selenium
  doped bismuth telluride.
\newblock \emph{Advanced Materials Interfaces}, 8\penalty0 (15):\penalty0
  2100185, 2021.

\bibitem[Oztan et~al.(2022)Oztan, Welch, and LeBlanc]{oztan2022additive}
Cagri Oztan, Ryan Welch, and Saniya LeBlanc.
\newblock Additive manufacturing of bulk thermoelectric architectures: a
  review.
\newblock \emph{Energies}, 15\penalty0 (9):\penalty0 3121, 2022.

\bibitem[Cain et~al.(2016)Cain, Hanson, Shi, and Dravid]{cain2016emerging}
Jeffrey~D Cain, Eve~D Hanson, Fengyuan Shi, and Vinayak~P Dravid.
\newblock Emerging opportunities in the two-dimensional chalcogenide systems
  and architecture.
\newblock \emph{Current Opinion in Solid State and Materials Science},
  20\penalty0 (6):\penalty0 374--387, 2016.

\bibitem[Romanenko et~al.(2021)Romanenko, Chebanova, Chen, Su, and
  Wang]{romanenko2021review}
AI~Romanenko, GE~Chebanova, Tingting Chen, Wenbin Su, and Hongchao Wang.
\newblock Review of the thermoelectric properties of layered oxides and
  chalcogenides.
\newblock \emph{Journal of Physics D: Applied Physics}, 55\penalty0
  (14):\penalty0 143001, 2021.

\bibitem[Pan et~al.(2015)Pan, Wei, Wu, and Li]{pan2015electrical}
Yu~Pan, Tian-Ran Wei, Chao-Feng Wu, and Jing-Feng Li.
\newblock Electrical and thermal transport properties of spark plasma sintered
  n-type bi 2 te 3- x se x alloys: the combined effect of point defect and se
  content.
\newblock \emph{Journal of Materials Chemistry C}, 3\penalty0 (40):\penalty0
  10583--10589, 2015.

\bibitem[Shen et~al.(2011)Shen, Hu, Zhu, and Zhao]{shen2011texture}
JJ~Shen, LP~Hu, TJ~Zhu, and XB~Zhao.
\newblock The texture related anisotropy of thermoelectric properties in
  bismuth telluride based polycrystalline alloys.
\newblock \emph{Applied Physics Letters}, 99\penalty0 (12), 2011.

\bibitem[Pan and Li(2016)]{pan2016thermoelectric}
Yu~Pan and Jing-Feng Li.
\newblock Thermoelectric performance enhancement in n-type bi2 (tese) 3 alloys
  owing to nanoscale inhomogeneity combined with a spark plasma-textured
  microstructure.
\newblock \emph{NPG Asia Materials}, 8\penalty0 (6):\penalty0 e275--e275, 2016.

\bibitem[Tang et~al.(2019)Tang, Huang, Pei, Zhang, Shang, Shan, Zhang, Gu, and
  Wen]{tang2019bi}
Chunmei Tang, Zhicheng Huang, Jun Pei, Bo-Ping Zhang, Peng-Peng Shang, Zhihang
  Shan, Zhiyue Zhang, Haiyun Gu, and Kaibin Wen.
\newblock Bi 2 te 3 single crystals with high room-temperature thermoelectric
  performance enhanced by manipulating point defects based on first-principles
  calculation.
\newblock \emph{RSC advances}, 9\penalty0 (25):\penalty0 14422--14431, 2019.

\bibitem[Freysoldt et~al.(2014)Freysoldt, Grabowski, Hickel, Neugebauer,
  et~al.]{RevModPhys.86.253}
Christoph Freysoldt, Blazej Grabowski, Tilmann Hickel, J\"org Neugebauer,
  et~al.
\newblock First-principles calculations for point defects in solids.
\newblock \emph{Rev. Mod. Phys.}, 86:\penalty0 253--305, Mar 2014.

\bibitem[Chun et~al.(2020)Chun, Roy, Nguyen, Choi, Udaykumar, and
  Baek]{chun2020deep}
Sehyun Chun, Sidhartha Roy, Yen~Thi Nguyen, Joseph~B Choi, Holavanahalli~S
  Udaykumar, and Stephen~S Baek.
\newblock Deep learning for synthetic microstructure generation in a
  materials-by-design framework for heterogeneous energetic materials.
\newblock \emph{Scientific reports}, 10\penalty0 (1):\penalty0 13307, 2020.

\bibitem[Dong et~al.(2024)Dong, Fu, Siriwardane, and Hu]{dong2024generative}
Rongzhi Dong, Nihang Fu, Edirisuriya~MD Siriwardane, and Jianjun Hu.
\newblock Generative design of inorganic compounds using deep diffusion
  language models.
\newblock \emph{The Journal of Physical Chemistry A}, 128\penalty0
  (29):\penalty0 5980--5989, 2024.

\bibitem[Park et~al.(2024{\natexlab{a}})Park, Gill, Moosavi, and
  Kim]{park2024inverse}
Junkil Park, Aseem Partap~Singh Gill, Seyed~Mohamad Moosavi, and Jihan Kim.
\newblock Inverse design of porous materials: a diffusion model approach.
\newblock \emph{Journal of Materials Chemistry A}, 12\penalty0 (11):\penalty0
  6507--6514, 2024{\natexlab{a}}.

\bibitem[Takahara et~al.(2024)Takahara, Shibata, and
  Mizoguchi]{takahara2024generative}
Izumi Takahara, Kiyou Shibata, and Teruyasu Mizoguchi.
\newblock Generative inverse design of crystal structures via diffusion models
  with transformers.
\newblock \emph{arXiv preprint arXiv:2406.09263}, 2024.

\bibitem[Christopher et~al.(2024)Christopher, Baek, and
  Fioretto]{christopher2024constrained}
Jacob~K Christopher, Stephen Baek, and Nando Fioretto.
\newblock Constrained synthesis with projected diffusion models.
\newblock \emph{Advances in Neural Information Processing Systems},
  37:\penalty0 89307--89333, 2024.

\bibitem[Cheng et~al.(2024)Cheng, Han, Maddix, Ansari, Stuart, Mahoney, and
  Wang]{cheng2024gradient}
Chaoran Cheng, Boran Han, Danielle~C Maddix, Abdul~Fatir Ansari, Andrew Stuart,
  Michael~W Mahoney, and Yuyang Wang.
\newblock Gradient-free generation for hard-constrained systems.
\newblock \emph{arXiv preprint arXiv:2412.01786}, 2024.

\bibitem[Utkarsh et~al.(2025)Utkarsh, Cai, Edelman, Gomez-Bombarelli, and
  Rackauckas]{utkarsh2025physics}
Utkarsh Utkarsh, Pengfei Cai, Alan Edelman, Rafael Gomez-Bombarelli, and
  Christopher~Vincent Rackauckas.
\newblock Physics-constrained flow matching: Sampling generative models with
  hard constraints.
\newblock \emph{arXiv preprint arXiv:2506.04171}, 2025.

\bibitem[Yuan et~al.(2023)Yuan, Song, Iqbal, Vahdat, and
  Kautz]{yuan2023physdiff}
Ye~Yuan, Jiaming Song, Umar Iqbal, Arash Vahdat, and Jan Kautz.
\newblock Physdiff: Physics-guided human motion diffusion model.
\newblock In \emph{Proceedings of the IEEE/CVF international conference on
  computer vision}, pages 16010--16021, 2023.

\bibitem[Zampini et~al.(2025)Zampini, Christopher, Oneto, Anguita, and
  Fioretto]{zampini2025training}
Stefano Zampini, Jacob~K Christopher, Luca Oneto, Davide Anguita, and
  Ferdinando Fioretto.
\newblock Training-free constrained generation with stable diffusion models.
\newblock \emph{arXiv preprint arXiv:2502.05625}, 2025.

\bibitem[Giannozzi et~al.(2009)Giannozzi, Baroni, Bonini, Calandra, Car,
  Cavazzoni, Ceresoli, Chiarotti, Cococcioni, Dabo,
  et~al.]{giannozzi2009quantum}
Paolo Giannozzi, Stefano Baroni, Nicola Bonini, Matteo Calandra, Roberto Car,
  Carlo Cavazzoni, Davide Ceresoli, Guido~L Chiarotti, Matteo Cococcioni,
  Ismaila Dabo, et~al.
\newblock Quantum espresso: a modular and open-source software project for
  quantumsimulations of materials.
\newblock \emph{Journal of physics: Condensed matter}, 21\penalty0
  (39):\penalty0 395502, 2009.

\bibitem[Song and Ermon(2019)]{song2019generative}
Yang Song and Stefano Ermon.
\newblock Generative modeling by estimating gradients of the data distribution.
\newblock \emph{Advances in neural information processing systems}, 32, 2019.

\bibitem[Song et~al.(2020)Song, Sohl-Dickstein, Kingma, Kumar, Ermon, and
  Poole]{song2020score}
Yang Song, Jascha Sohl-Dickstein, Diederik~P Kingma, Abhishek Kumar, Stefano
  Ermon, and Ben Poole.
\newblock Score-based generative modeling through stochastic differential
  equations.
\newblock \emph{arXiv preprint arXiv:2011.13456}, 2020.

\bibitem[Ho et~al.(2020)Ho, Jain, and Abbeel]{ho2020denoising}
Jonathan Ho, Ajay Jain, and Pieter Abbeel.
\newblock Denoising diffusion probabilistic models.
\newblock \emph{Advances in neural information processing systems},
  33:\penalty0 6840--6851, 2020.

\bibitem[Park et~al.(2024{\natexlab{b}})Park, McCann, Garcia-Cardona, Wohlberg,
  and Kamilov]{park2024random}
Chicago~Y Park, Michael~T McCann, Cristina Garcia-Cardona, Brendt Wohlberg, and
  Ulugbek~S Kamilov.
\newblock Random walks with tweedie: A unified framework for diffusion models.
\newblock \emph{arXiv preprint arXiv:2411.18702}, 2024{\natexlab{b}}.

\bibitem[Batatia et~al.(2022)Batatia, Kovacs, Simm, Ortner, and
  Cs{\'a}nyi]{batatia2022mace}
Ilyes Batatia, David~P Kovacs, Gregor Simm, Christoph Ortner, and G{\'a}bor
  Cs{\'a}nyi.
\newblock Mace: Higher order equivariant message passing neural networks for
  fast and accurate force fields.
\newblock \emph{Advances in neural information processing systems},
  35:\penalty0 11423--11436, 2022.

\bibitem[Liang et~al.(2025)Liang, Christopher, Koenig, and
  Fioretto]{liang2025simultaneous}
Jinhao Liang, Jacob~K Christopher, Sven Koenig, and Ferdinando Fioretto.
\newblock Simultaneous multi-robot motion planning with projected diffusion
  models.
\newblock \emph{arXiv preprint arXiv:2502.03607}, 2025.

\bibitem[Ye et~al.(2024)Ye, Lin, Han, Xu, Liu, Liang, Ma, Zou, and
  Ermon]{ye2024tfg}
Haotian Ye, Haowei Lin, Jiaqi Han, Minkai Xu, Sheng Liu, Yitao Liang, Jianzhu
  Ma, James~Y Zou, and Stefano Ermon.
\newblock Tfg: Unified training-free guidance for diffusion models.
\newblock \emph{Advances in Neural Information Processing Systems},
  37:\penalty0 22370--22417, 2024.

\bibitem[Yu et~al.(2023)Yu, Wang, Zhao, Ghanem, and Zhang]{yu2023freedom}
Jiwen Yu, Yinhuai Wang, Chen Zhao, Bernard Ghanem, and Jian Zhang.
\newblock Freedom: Training-free energy-guided conditional diffusion model.
\newblock In \emph{Proceedings of the IEEE/CVF International Conference on
  Computer Vision}, pages 23174--23184, 2023.

\bibitem[Cornwall et~al.(2024)Cornwall, Meyers, Day, Wollman, Dalchau, and
  Sim]{cornwalltraining}
Lewis Cornwall, Joshua Meyers, James Day, Lilly~S Wollman, Neil Dalchau, and
  Aaron Sim.
\newblock Training-free guidance of diffusion models for generalised
  inpainting, 2024.

\bibitem[Shen et~al.(2024)Shen, Jiang, Yang, Wang, Han, and
  Li]{shen2024understanding}
Yifei Shen, Xinyang Jiang, Yifan Yang, Yezhen Wang, Dongqi Han, and Dongsheng
  Li.
\newblock Understanding and improving training-free loss-based diffusion
  guidance.
\newblock \emph{Advances in Neural Information Processing Systems},
  37:\penalty0 108974--109002, 2024.

\bibitem[Vaeth et~al.(2024)Vaeth, Fruehwald, Paassen, and
  Gregorova]{vaeth2024gradcheck}
Philipp Vaeth, Alexander~M Fruehwald, Benjamin Paassen, and Magda Gregorova.
\newblock Gradcheck: Analyzing classifier guidance gradients for conditional
  diffusion sampling.
\newblock \emph{arXiv preprint arXiv:2406.17399}, 2024.

\bibitem[Kawar et~al.(2022)Kawar, Ganz, and Elad]{kawar2022enhancing}
Bahjat Kawar, Roy Ganz, and Michael Elad.
\newblock Enhancing diffusion-based image synthesis with robust classifier
  guidance.
\newblock \emph{arXiv preprint arXiv:2208.08664}, 2022.

\bibitem[Rawlings et~al.(2020)Rawlings, Mayne, Diehl,
  et~al.]{rawlings2020model}
James~Blake Rawlings, David~Q Mayne, Moritz Diehl, et~al.
\newblock \emph{Model predictive control: theory, computation, and design},
  volume~2.
\newblock Nob Hill Publishing Madison, WI, 2020.

\bibitem[Boyd and Vandenberghe(2004)]{boyd2004convex}
Stephen~P Boyd and Lieven Vandenberghe.
\newblock \emph{Convex optimization}.
\newblock Cambridge university press, 2004.

\bibitem[Fioretto et~al.(2020)Fioretto, Van~Hentenryck, Mak, Tran, Baldo, and
  Lombardi]{fioretto2020lagrangian}
Ferdinando Fioretto, Pascal Van~Hentenryck, Terrence~WK Mak, Cuong Tran,
  Federico Baldo, and Michele Lombardi.
\newblock Lagrangian duality for constrained deep learning.
\newblock In \emph{Joint European conference on machine learning and knowledge
  discovery in databases}, pages 118--135. Springer, 2020.

\bibitem[Ho and Salimans(2022)]{ho2022classifier}
Jonathan Ho and Tim Salimans.
\newblock Classifier-free diffusion guidance.
\newblock \emph{arXiv preprint arXiv:2207.12598}, 2022.

\bibitem[Giannone et~al.(2023)Giannone, Srivastava, Winther, and
  Ahmed]{giannone2023aligning}
Giorgio Giannone, Akash Srivastava, Ole Winther, and Faez Ahmed.
\newblock Aligning optimization trajectories with diffusion models for
  constrained design generation.
\newblock \emph{Advances in neural information processing systems},
  36:\penalty0 51830--51861, 2023.

\bibitem[Dhariwal and Nichol(2021)]{dhariwal2021diffusion}
Prafulla Dhariwal and Alexander Nichol.
\newblock Diffusion models beat gans on image synthesis.
\newblock \emph{Advances in neural information processing systems},
  34:\penalty0 8780--8794, 2021.

\bibitem[Hashibon and Els\"asser(2011)]{PhysRevB.84.144117}
Adham Hashibon and Christian Els\"asser.
\newblock First-principles density functional theory study of native point
  defects in bi${}_{2}$te${}_{3}$.
\newblock \emph{Phys. Rev. B}, 84:\penalty0 144117, Oct 2011.
\newblock \doi{10.1103/PhysRevB.84.144117}.

\bibitem[Virtanen and {others}(2020)]{virtanen2020scipy}
Pauli Virtanen and {others}.
\newblock {SciPy} 1.0: Fundamental algorithms for scientific computing in
  python.
\newblock \emph{Nature Methods}, 17:\penalty0 261--272, 2020.

\bibitem[Yildirim and Brown(2021)]{rdfpy}
Batuhan Yildirim and Hamish~Galloway Brown.
\newblock by256/rdfpy: rdfpy-v1.0.0, March 2021.
\newblock URL \url{https://doi.org/10.5281/zenodo.4625675}.

\end{thebibliography}
\bibliographystyle{unsrtnat}

\appendix

\section{Related Work}
\label{app:related_work}

\paragraph{Constraint Conditioning} \cite{dhariwal2021diffusion, ho2022classifier}
enables controllable generation by incorporating a conditioning variable to bias the posterior which is sampled. Accomplished either explicitly through the addition of classifier-derived gradient signals or implicitly by training the denoiser with specific conditioning labels, these techniques provide soft guidance towards particular subdistributions.
Notably, these methods have been applied to improve constraint adherence in a variety of applications including [fill in references].
Yet, while constraint conditioning can improve feasibility rates in specific cases, it is unreliable when exact satisfaction is required. Particularly when constraint sets are complex, conditioning methods have been shown ineffective in providing viable outputs, a challenge that is demonstrated in our empirical analysis.

\paragraph{Post-Processing Optimization} \cite{giannone2023aligning}
provides an alternative approach which injects constraints following the denoising process. These approaches leverage a generative model to produce a starting structure \(\bm{x}_0\), after which domain-specific corrections are applied. For material structure generations, this often consists of running DFT simulators \cite{giannozzi2009quantum} to relax the atomic configurations, ensuring stability as the structure is refined to reach equilibrium.
Yet, these approaches are inherently void of distributional information, and the optimization procedure may drive the samples away from the learned distribution. While the original candidate structure will fall within \(p_\text{data}\), without access to the learned score function or likelihood estimates, post-processing can degrade sample quality, resulting in an output distribution that is constrained but no longer resembles the training data.

\section{Augmented Lagrangian Method}
\label{app:alm}

In our settings, projection is required to ensure a collection of physical constraints. Unlike standard convex projection operators, these constraints are often highly non-convex and some are evaluated through surrogate models, such as pretrained MACE model. This makes exact projection onto feasible set intractable. To overcome this, we rely on the Augmented Lagrangian Method (ALM) \cite{boyd2004convex}. Instead of attempting to solve the constrained problem in a closed form, Lagrangian relaxation converts these into differentiable residuals. By combining linear multipliers $\lambda$ and quadratic penalties $\mu$, ALM provides a mechanism to gradually enforce feasibility. 

Let $\bm{x}_0$ denote the current sample produced by the diffusion process, and let $\bm{y}$ be the projected candidate. 
We aim to find $\bm{y}$ that remains close to $\bm{x}_t$ while reducing constraint violations. The approach introduces a relaxed objective by embedding constraint residuals into the optimization problem:

\begin{align}
    \mathcal{L}(\bm{y},\bm{x}_0^{(i)}; \lambda, \mu) := \|\bm{y} - \bm{x}_0^{(i)}\|^2_2 + w \cdot \|\mathrm{force}(\bm{y})\|_2^2 + \sum_j^N \lambda_j c_j(\bm{y}) + \sum_k^N \frac{\mu_k}{2} c_k(\bm{y})^2
\end{align}

where \(\{c_1, \ldots, c_N\}\) denotes a series of differentiable violation residual of constraints. Additionally, we treat the \(\mathrm{force}\) prediction provided by MACE \cite{batatia2022mace} as a minimization term (scaled by \(w\)) rather than as a hard constraint,
due to runtime considerations and surrogate predictive accuracy.  This formulation transforms the constrained projection into a differentiable optimization problem that can be solved iteratively alongside diffusion sampling. Algorithm \ref{algorithm_alm} summarizes the procedure: starting from
the diffusion output $\bm{x}_t$, we iteratively compute constraint residuals, update the augmented objective, take a gradient step, and adjust multipliers and penalties.

\begin{wrapfigure}[10]{r}{0.55\linewidth}
    \vspace{-15pt}
    \begin{algorithm}[H]
            \DontPrintSemicolon
            \LinesNotNumbered
            \caption{\footnotesize Augmented Lagrangian Projection}
            \label{algorithm_alm}
            {\footnotesize
            \KwIn{$\bm{x}_t$, Lagrange multiplier: $\lambda$, quadratic penalty: $\mu$, scaling constant: $\alpha$, step size: $\gamma$, tolerance: $\delta$}
            
            $\bm{y} \gets \bm{x}_t$
            
            \While{$\sum_i c_i(\bm{y}) < \delta$}{
                \For{$j \gets 1$ \KwTo max\_inner\_iter}{
                    
                    $\mathcal{L}_{\text{ALM}} \gets \|\bm{x}_t - \bm{y}\|_2^2 + \sum_j^N \lambda_j c_j(\bm{y}) + \sum_k^N \frac{\mu_k}{2} c_k(\bm{y})^2$
                    
                    $\bm{y} \gets \bm{y} - \gamma\, \nabla_{\bm{y}} \mathcal{L}_{\text{ALM}}$
                    
                }
                $\lambda \gets \lambda + \mu\, \sum_j^N \lambda_j c_j(\bm{y}); \;\;
                \mu \gets \min\bigl(\alpha\mu,\, \mu_{\text{max}}\bigr)$\;
                
            }
            $\bm{x}_{t-\Delta} \gets \bm{y}$ 
            
            \Return $\bm{x}_{t-\Delta}$\;
            }
        \end{algorithm}
\end{wrapfigure}

In our implementation, the residual vector $\tilde{\phi}(y)$ collects violations from three major classes of constraints: (i) geometric constraints, including minimum interatomic distances and periodic boundaries; (ii) distributional constraints, capturing radial distribution alignment; and (iii) force minimization, based on surrogate predicted forces.

\section{Evaluation Setup}
\label{app:setup}

Our model is trained on full DFT relaxation trajectories. The training dataset consists of approximately 13,000 structures from DFT trajectories. Each structure is represented by atom types and corresponding 3D coordinates in Cartesian space. Our generative model starts from Gaussian noise and progressively denoises toward physically relaxed structures under the targeted defect configurations. Each of the six stoichiometric configurations was evaluated with 100 generated samples, resulting in 600 structures per method. These six configurations were chosen because they are the most prevalent in our training trajectories and serve as representative cases of the four major defect categories considered \cite{PhysRevB.84.144117}. All methods are evaluated under a shared random seeds to ensure comparability. 

\paragraph{Metric computation}
For pairwise RMSD similarity, given a generated structure and the set of relaxed references under the same stoichiometric composition. Before measuring, both structures are centered, and atoms are matched by species. Within each species, we form the pairwise Eucluidean distance matrix between generated and reference coordinates and solve an optimal one to one assignment using the Hungarian algorithm\cite{virtanen2020scipy}. The final RMSD is taken under this assignment and reported at the structure level of the sampling set.

The RDF characterizes the radial number density of neighbors, i.e., the probability density of finding another atom at distance $r$\cite{rdfpy}. We compute $g(r)$ for generated structures under a 5\r{A} local cutoff. Among relaxed references of the same composition, we then identify the nearest reference in RDF space by minimizing the RMSD between RDF profiles. The result is recorded as the sample’s distributional deviation. This procedure provides a global distribution check that is independent of the previous RMSD assignment.

To assess proximity to locally stable states, we estimate total forces for all generated samples with a pretrained surrogate and summarize each structure by the magnitude of its total forces. Lower values indicate that the generated configuration lies closer to an equilibrium basin and is more suitable as an input to downstream relaxation. We report mean total forces over the sampled set to enable comparison across methods.

\begin{table}[H]
\centering
\label{tab:methods_ex}
\begin{tabular}{llccc}
\toprule
Method & Config & RMSD $\downarrow$ & RDF $\downarrow$ & Force (eV/\r{A}) $\downarrow$  \\
\midrule
\multirow{6}{*}{Conditional DM} 
 & 16Bi + 21Te & $1.86 \pm 0.33$ & $65.85 \pm 19.75$ & $1.85 \times 10^{9}$ \\
 & 16Bi + 22Te & $1.93 \pm 0.36$ & $82.11 \pm 29.54$ & $3.49 \times 10^{9}$ \\
 & 18Bi + 22Te & $2.10 \pm 0.38$ & $87.90 \pm 39.29$ & $1.64 \times 10^{9}$ \\
 & 14Bi + 26Te & $1.22 \pm 0.60$ & $40.63 \pm 41.77$ & $1.28 \times 10^{9}$ \\
 & 14Bi + 24Te & $0.75 \pm 0.81$ & $27.41 \pm 44.46$ & $5.52 \times 10^{8}$ \\
 & 13Bi + 24Te & $0.95 \pm 0.51$ & $22.17 \pm 33.53$ & $3.90 \times 10^{8}$ \\
\midrule
\multirow{6}{*}{Projected DM} 
 & 16Bi + 21Te & $2.54 \pm 0.55$ & $1.11 \pm 0.77$ & $8.35 \times 10^{2}$ \\
 & 16Bi + 22Te & $1.41 \pm 0.83$ & $0.78 \pm 0.55$ & $7.20 \times 10^{1}$ \\
 & 18Bi + 22Te & $1.85 \pm 1.17$ & $1.45 \pm 1.62$ & $5.85 \times 10^{3}$ \\
 & 14Bi + 26Te & $1.43 \pm 0.78$ & $1.00 \pm 0.67$ & $5.04 \times 10^{1}$ \\
 & 14Bi + 24Te & $2.35 \pm 0.45$ & $0.94 \pm 0.49$ & $5.50 \times 10^{1}$ \\
 & 13Bi + 24Te & $2.93 \pm 0.39$ & $1.44 \pm 0.38$ & $1.17 \times 10^{2}$ \\
\midrule
\multirow{6}{*}{Post-proc DM} 
 & 16Bi + 21Te & $2.30 \pm 0.40$ & $27.25 \pm 28.30$ & $8.45 \times 10^{6}$ \\
 & 16Bi + 22Te & $1.14 \pm 0.65$ & $21.82 \pm 28.09$ & $9.29 \times 10^{6}$ \\
 & 18Bi + 22Te & $1.41 \pm 0.45$ & $63.27 \pm 35.58$ & $1.69 \times 10^{8}$ \\
 & 14Bi + 26Te & $1.16 \pm 0.62$ & $50.00 \pm 35.32$ & $6.54 \times 10^{7}$ \\
 & 14Bi + 24Te & $2.17 \pm 0.38$ & $37.37 \pm 28.29$ & $1.61 \times 10^{7}$ \\
 & 13Bi + 24Te & $2.75 \pm 0.34$ & $59.24 \pm 28.05$ & $1.76 \times 10^{7}$ \\
\midrule
\multirow{6}{*}{\textbf{Ours}} 
 & 16Bi + 21Te & $0.90 \pm 0.38$ & $0.30 \pm 0.11$ & $7.88 \times 10^{-2}$ \\
 & 16Bi + 22Te & $1.02 \pm 0.70$ & $0.35 \pm 0.18$ & $1.44 \times 10^{-1}$ \\
 & 18Bi + 22Te & $1.22 \pm 0.42$ & $0.47 \pm 0.22$ & $3.00 \times 10^{-1}$ \\
 & 14Bi + 26Te & $1.00 \pm 0.39$ & $0.28 \pm 0.19$ & $1.67 \times 10^{-1}$ \\
 & 14Bi + 24Te & $0.68 \pm 0.56$ & $0.30 \pm 0.11$ & $1.80 \times 10^{-1}$ \\
 & 13Bi + 24Te & $0.99 \pm 0.49$ & $0.36 \pm 0.06$ & $1.93 \times 10^{-1}$ \\
\bottomrule
\end{tabular}
\vspace{10pt}
\caption{Complete experimental results across all six stoichiometric configurations in four defect categories: Te vacancies (16Bi + 22Te, 16Bi + 21Te), Bi vacancies (14Bi + 24Te, 13Bi + 24Te), Bi in Te antisites (18Bi + 22Te), and Te in Bi antisites (14Bi + 26Te).}
\end{table}

\end{document}